\documentstyle[aps,preprint]{revtex}

\begin{document}


\title{THE EFFECTIVE BOSONIC HAMILTONIAN FOR EXCITONS RECONSIDERED}
\author{M. Combescot and O. Betbeder-Matibet}
\address{Groupe de Physique des Solides, Universit\'{e} Denis Diderot
et
Universit\'{e} Pierre et Marie Curie,\\
CNRS, Tour 23, 2 place Jussieu, 75251 Paris Cedex 05, France}
\maketitle
\begin{abstract}
The effective bosonic hamiltonian for excitons, extensively quoted up
to
now, cannot be correct because it is (surprisingly) non-hermitian.
The
oversight physically originates from the intrinsic difficulty of
properly
defining electron-hole interactions {\em between} excitons when
dealing
with {\em exchange} terms. By using our commutation technique, we
show
that the fermionic character of the excitons cannot be forced into a
dressed Coulomb interaction only : The effective bosonic hamiltonian
must
contain purely fermionic terms of the same order as the Coulomb
terms.
They are necessary to ensure hermiticity, and they do not reduce to a
two-body interaction, Pauli exclusion being N-body by essence.
\end{abstract}

\pacs{PACS numbers: }


In the very low density and temperature regime, the electron-hole
(e-h)
pairs of a semiconductor are bound into hydrogen-like states known as
excitons. In this regime, it is widely accepted
\cite{Hanamura,Haug,Ivanov,Axt,Ciuti,Inoue} that the semiconductor
can be
represented by an effective hamiltonian,
\begin{equation}
H_{\rm \em eff} = \sum_{i} E_{i} \, \overline{B}_{i}^{+} \,
\overline{B}_{i} +
\frac{1}{2} \, \sum_{ijmn} {\cal E}_{mnij} \, \overline{B}_{m}^{+} \,
\overline{B}_{n}^{+} \, \overline{B}_{i} \, \overline{B}_{j}  ,
\label{eq1}
\end{equation}
in which the excitons are assumed to be real bosons :
$\overline{B}_{i}^{+}$ is the $i$ boson-exciton creation operator,
$[\overline{B}_{i},\overline{B}_{j}^{+}] = \delta_{ij}$, $i$ standing
for
($\nu_{i},{\mathbf q}_{i}$), where $\nu_{i}$ characterizes the
exciton
level and ${\mathbf q}_{i}$ its center of mass momentum, the exciton
energy being $E_{i} = \varepsilon_{\nu_{i}}+E_{{\mathbf q}_{i}}$. The
second term of Eq.~(\ref{eq1}) comes from interactions {\em between}
excitons. The $X$-$X$ scattering ${\cal E}_{mnij}$, reported up to
now, is
a sum of a direct and an exchange Coulomb term. This exchange term
results
from the composite nature of the excitons. There are in fact two ways
of
forming two excitons with two electrons ($e_{1},e_{2}$) and two holes
($h_{1},h_{2}$), depending on how the e and h are coupled. This quite
harmless evidence generates major difficulties when dealing with e-h
Coulomb interactions. Indeed, $V({\mathbf r}_{e_1}-{\mathbf r}_{h_2})
=
V_{e_1 h_2}$ is an interaction {\em between} two excitons if they are
formed with ($e_{1},h_{1}$) and ($e_{2},h_{2}$), while it is an
interaction {\em inside} one of them if they are formed with
($e_{1},h_{2}$) and ($e_{2},h_{1}$). The concept of e-h interaction
{\em
between} excitons is thus ambiguous for exchange processes. One of
the
problems of the effective hamiltonian used up to now comes from this
difficulty.

In this letter, we first reconsider the widely accepted exciton
effective
hamiltonian, and point out its (surprising) non hermiticity. In a
second
part, we recall the commutation technique introduced in a previous
problem
\cite{MCRC,MCReport} dealing with interacting excitons, namely the
exciton
optical Stark shift, and we use it to derive a new $X$-$X$
scattering. It
contains a purely fermionic contribution, necessary to ensure
hermiticity,
which is conceptually quite different from a Coulomb interaction
dressed
by fermionic effects. We also find that for $N>2$ excitons, the
effective
bosonic hamiltonian must contain (3,...,$N$)-body scatterings of the
same
order as the Coulomb terms. They have a purely fermionic origin :
While Coulomb interaction is basically 2-body, Pauli exclusion
between the
electrons of the $N$ excitons is intrinsically $N$-body, so that one
cannot get rid of it by 2-body operators only.

\textbf{1. The former result.~} The former ${\cal E}_{mnij}$ reads
\cite{note9}
\begin{eqnarray}
{\cal E}_{mnij} & = & \int de_1 de_2 dh_1 dh_2 \, \phi_m^*(e_1,h_1)
\,
\phi_n^*(e_2,h_2) \, \big( V_{e_1 e_2} + V_{h_1 h_2} -V_{e_1
h_2}-V_{e_2
h_1} \big) \nonumber\\
 & & \hskip2cm \times [\phi_i(e_1,h_1) \,
\phi_j(e_2,h_2)-\phi_i(e_1,h_2)
\, \phi_j(e_2,h_1)] ,
\label{eq2}
\end{eqnarray}
where $e$ stands for ${\mathbf r}_e$ and $\phi_i(e,h)$ is the wave
function of the $i$ exciton. The first term of the bracket gives the
direct part ${\cal E}_{mnij}^{\rm dir}$ of the $X$-$X$ interaction,
while
the usual exchange part ${\cal E}_{mnij}^{\rm exc}$ comes from the
second
term. Although widely quoted, this $X$-$X$ scattering cannot be
correct:
Indeed, as ${\cal E}_{mnij} \neq {\cal E}_{ijmn}^*$, the
corresponding
effective hamiltonian is non-hermitian. Strangely enough, this
alarming
point seems to have stayed unnoticed up to now. The problem comes
from the
e-h interactions in the exchange part. These e-h interactions a
priori
contain four terms, $V_{e_1 h_1} + V_{e_2 h_2} + V_{e_1 h_2} + V_{e_2
h_1}$.  $V_{e_1 h_1}$ (resp. $V_{e_2 h_2}$) has clearly to be dropped
because it is a Coulomb interaction {\em inside} the $m$ (resp. $n$)
exciton. However, on this basis, $V_{e_1 h_2}$ (resp. $V_{e_2 h_1}$)
should be dropped from the exchange term because it is a Coulomb
interaction inside the $i$ (resp. $j$) ``exchange'' exciton. The
correct
$X$-$X$ exchange scattering should thus contain either no e-h
interaction
at all, or possibly all them four : There is no reason to keep two of
them
only \cite{note9}.

\vskip5mm

\textbf{2. The commutation technique.~} The e-h interaction {\em
between}
excitons is in
fact quite subtle, as it is impossible to split the e-h terms of the
bare
Coulomb interaction operator into a part which binds the exciton
(through
repeated e-h interactions) and a part which makes the excitons to
interact.

In one of our previous works, we already faced such a difficulty. We
overcame it by introducing \cite{MCRC,MCReport} the operator $V_i^+$
defined as $[H_{sc},B_i^+] = E_i \, B_i^+ + V_i^+$, where $H_{sc}$ is
the
{\em exact} semiconductor hamiltonian, and $B_i^+$ the {\em exact}
exciton
creation operator given in the appendix (Eq.~(\ref{eqa1})). If the
excitons were non-interacting, we would have $[H_{sc},B_i^+] = E_i \,
B_i^+$ only, so that $V_i^+$ does come from
interaction {\em between} excitons. From its explicit value given in
Eq.~(\ref{eqa2}), we get
\begin{equation}
[[H_{sc},B_i^+],B_j^+] = [V_i^+,B_j^+] = \sum_{m n} \xi_{mnij}^{\rm
dir}
\, B_m^+ \, B_n^+ ,
\label{eq3}
\end{equation}
where $\xi_{mnij}^{\rm dir} = (\xi_{ijmn}^{\rm dir})^*$ is just
${\cal
E}_{mnij}^{\rm dir}$ properly symmetrized (see Eq.~(\ref{eqa6})).

Besides $V_i^+$, there is another ``interaction'' between excitons
which
plays a crucial part. It comes
from the fermionic character of the exciton which appears through the
boson-departure operator $D_{i j} = \delta_{i j}-[B_i,B_j^+]$. When
acting
on $B_j^+$, this operator gives
\begin{equation}
[D_{m i},B_j^+] = - [[B_m,B_i^+],B_j^+] = 2 \sum_n \lambda_{mnij} \,
B_n^+ ,
\label{eq4}
\end{equation}
with $\lambda_{mnij} = \lambda_{ijmn}^*$ given in Eq.~(\ref{eqa8}).
$\lambda_{mnij}$ corresponds to cross the e (or the h) of two
excitons, so that it relates two excitons with e and h bound in a
different way:
\begin{equation}
B_i^+ \, B_j^+ = - \sum_{m n} \lambda_{mnij} \, B_m^+ \, B_n^+ .
\label{eq5}
\end{equation}
Equations (\ref{eq3},\ref{eq4}) are the key equations of our
commutation
technique. They allow to calculate any quantity dealing with
interacting
excitons.

\vskip5mm

\textbf{3. The two exchange terms.~} From them we get
\begin{eqnarray}
H_{sc} \, B_i^+ \, B_j^+ | 0 \rangle & = & (E_i + E_j) \, B_i^+ \,
B_j^+ |
0 \rangle + \sum_{m n} \xi_{mnij}^{\rm dir} \, B_m^+ \, B_n^+ | 0
\rangle
, \label{eq6} \\
\langle 0 | B_m \, B_n \, B_i^+ \, B_j^+ | 0 \rangle & = & \delta_{m
i} \,
\delta_{n j} + \delta_{m j} \, \delta_{n i} - 2 \, \lambda_{mnij}.
\label{eq7}
\end{eqnarray}
This gives for the matrix elements of the exact hamiltonian $H_{sc}$,
calculated with $H_{sc}$ acting on the right or on the left,
\begin{eqnarray}
\langle 0 | B_m \, B_n \, H_{sc} \, B_i^+ \, B_j^+ | 0 \rangle & = &
(E_i
+ E_j) \, (\delta_{m i} \, \delta_{n j} + \delta_{m j} \, \delta_{n
i} - 2
\, \lambda_{mnij}) + 2 \, (\xi_{mnij}^{\rm dir} - \sum_{r s}
\lambda_{mnrs} \, \xi_{rsij}^{\rm dir})
\label{eq8} \\
& = & (E_m + E_n) \, (\delta_{m i} \, \delta_{n j} + \delta_{m j} \,
\delta_{n i} - 2 \, \lambda_{mnij}) + 2 \, (\xi_{mnij}^{\rm dir} -
\sum_{r
s} \xi_{mnrs}^{\rm dir} \, \lambda_{rsij}) . \label{eq9}
\end{eqnarray}
Two Coulomb exchange terms thus appear,
\begin{equation}
\xi_{mnij}^{\rm left} = \sum_{r s} \lambda_{mnrs} \, \xi_{rsij}^{\rm
dir},
\hskip 1cm \xi_{mnij}^{\rm right} = \sum_{r s} \xi_{mnrs}^{\rm dir}
\,
\lambda_{rsij} = (\xi_{ijmn}^{\rm left})^* .
\label{eq10}
\end{equation}
Due to Eqs.~(\ref{eq8}--\ref{eq9}),
they verify
\begin{equation}
(E_m + E_n) \, \lambda_{mnij} + \xi_{mnij}^{\rm right} = (E_i + E_j)
\,
\lambda_{mnij} + \xi_{mnij}^{\rm left} .
\label{eq11}
\end{equation}
They are thus equal for $E_m + E_n = E_i + E_j$ only, i.e. diagonal
processes, $(m n) = (i j)$, or possibly scattering between excitons
staying inside the same exciton level {\em and} having an infinite
total
mass. Using the expression of $\xi_{mnij}^{\rm right}$ given in the
appendix, we find that $\xi_{mnij}^{\rm right} = {\cal E}_{mnij}^{\rm
exc}$ (properly symmetrized), with its e-h interactions ``inside''
the $i$
and $j$ excitons. The other exchange term $\xi_{mnij}^{\rm left}$,
with
its e-h interactions ``inside'' the $m$ and $n$ excitons, does not
appear
\cite{Ciuti} in ${\cal E}_{mnij}$.

From Eqs.~(\ref{eqa4},\ref{eqa7}), we find that the $\lambda$'s as
well as
the direct, left and right $\xi$'s write as a sum over one free
${\mathbf
k}$ of a product of four $\langle {\mathbf k} | x_{\nu_i} \rangle$.
For
bound states, each $\langle {\mathbf k} | x_{\nu_i} \rangle$ induces
a
$(a_x^3/{\cal V})^{1/2}$ factor (where $a_x$ is the exciton Bohr
radius
and $\cal V$ the sample volume), so that the $\lambda$'s and the
various
$\xi$'s are {\em all} of the order of  $a_x^3/{\cal V}$ (even if
$\xi^{\rm
right}$ appears as a sum of $\xi^{\rm dir} \, \lambda$).

\vskip5mm

\textbf{4. The new $H_{\rm \em eff}.$~} We can think of identifying
$H_{\rm \em eff} \, \overline{B}_{i}^{+} \, \overline{B}_{j}^{+} \, |
0
\rangle$ with Eq.~(\ref{eq6}). This would lead to ${\cal E}_{mnij} =
\xi_{mnij}^{\rm dir}$. (Note that due to Eq.~(\ref{eq5}), we could
replace
the prefactor of the $B_m^+ \, B_n^+$ term in Eq.~(\ref{eq6}) by
$(a \, \xi_{mnij}^{\rm dir} - b \, \xi_{mnij}^{\rm left})/(a+b)$,
with
arbitrary ($a$, $b$); the hermiticity of $H_{\rm \em eff}$ however
forces
$b = 0$). As this $X$-$X$ scattering, without exchange terms,
completely
misses the fermionic character of the excitons, it looks reasonable
to
reject it.

A better way to determine an exciton effective bosonic hamiltonian is
to
force its matrix elements to be the same as the ones of the exact
hamiltonian. If we introduce the (normalized) $N$ exact-exciton state,
\begin{equation}
| \psi_{i_1, \cdots , i_N}^{(N)} \rangle = B_{i_1}^+ \, \cdots \,
B_{i_N}^+ | 0 \rangle/\langle 0 | B_{i_N} \, \cdots \, B_{i_1} \,
B_{i_1}^+ \, \cdots \, B_{i_N}^+ | 0 \rangle^{1/2} ,
\label{eq12}
\end{equation}
and the $N$ boson-exciton state $| \overline{\psi}_{i_1, \cdots ,
i_N}^{(N)} \rangle$  with $B_i^+$ replaced by $\overline{B}_i^+$,
this
condition reads
\begin{equation}
\langle \overline{\psi}_{i_1, \cdots , i_N}^{(N)} | H_{\rm \em eff} |
\overline{\psi}_{{i'}_1, \cdots , {i'}_N}^{(N)} \rangle = \langle
{\psi}_{i_1, \cdots , i_N}^{(N)} | H_{sc} | {\psi}_{{i'}_1, \cdots ,
{i'}_N}^{(N)} \rangle .
\label{eq13}
\end{equation}

i) In the {\em one-exciton} subspace, Eq.~(\ref{eq13}) immediately
gives
${\cal E}_{ij} = E_i \, \delta_{i j}$, for the one-body part $H_{\rm
\em
eff}$ written as $\displaystyle \sum_{i j} {\cal E}_{ij} \,
\overline{B}_{i}^{+} \, \overline{B}_{j}^{+}$.

ii) In the {\em two-exciton} subspace,
Eqs.~(\ref{eq1},\ref{eq7},\ref{eq9},\ref{eq13}) lead to
\begin{eqnarray}
{\cal E}_{mnij}^{\rm new} & = & \rho_{mnij} \, [\xi_{mnij}^{\rm dir}
-
\xi_{mnij}^{\rm right} - (1-\delta_{mnij}) \, (E_m + E_n) \,
\lambda_{mnij} ] \label{eq14} \\
\rho_{mnij} & = & \left[ \frac{(1+\delta_{mn}) \,
(1+\delta_{ij})}{(1+\delta_{mn}-2 \, \lambda_{mnmn}) \,
(1+\delta_{ij}- 2
\, \lambda_{ijij})} \right]^{1/2} , \label{eq15}
\end{eqnarray}
where $\delta_{mnij}=1$ for ($m n$)=($i j$) and 0 otherwise.
Eq.~(\ref{eq14}) shows that the non-diagonal scatterings have a
purely
fermionic contribution in $\lambda_{mnij}$, which is necessary to
ensure
the hermiticity of $H_{\rm \em eff}$ (see
Eqs.~(\ref{eq10},\ref{eq11})).
This ${\cal E}_{mnij}^{\rm new}$ can be rewritten in a more
symmetrical
way by introducing
\begin{eqnarray}
\xi_{mnij}^{\rm exc} & = & \frac{1}{2} \, (\xi_{mnij}^{\rm
right}+\xi_{mnij}^{\rm left}) = \frac{1}{2} \, \int de_1 de_2 dh_1
dh_2 \,
\phi_m^*(e_1,h_1) \, \phi_n^*(e_2,h_2)  \nonumber\\
 & & \times \big[ V_{e_1 e_2} + V_{h_1 h_2} - \frac{1}{2} \, (V_{e_1
h_1}
+ V_{e_2 h_2} + V_{e_1 h_2} + V_{e_2 h_1}) \big]  \, \phi_i(e_1,h_2)
\,
\phi_j(e_2,h_1) + (m \leftrightarrow n) ,
\label{eq16}
\end{eqnarray}
which is such that $\xi_{mnij}^{\rm exc} = (\xi_{ijmn}^{\rm exc})^*$.
From
Eqs.~(\ref{eq11},\ref{eq14},\ref{eq16}) we then get
\begin{eqnarray}
{\cal E}_{mnij}^{\rm new} & = & \xi_{mnij}^{\rm dir} -
\xi_{mnij}^{\rm
exc} - \eta_{mnij} + {\rm O}\big( (a_x^3/{\cal V})^{2} \big) ,
\label{eq17} \\
\eta_{mnij} & = & \frac{1}{2} \, (1 - \delta_{mnij}) \, (E_m +
E_n+E_i +
E_j) \, \lambda_{mnij} , \label{eq18}
\end{eqnarray}
as, for bound states, the $\lambda$ and the $\xi$'s are all in
$a_x^3/{\cal V}$, while $\rho_{mnij} = 1 + {\rm O}( a_x^3/{\cal V})$.

Equation~(\ref{eq17}) shows that the $a_x^3/{\cal V}$ leading term of
the
two-body part of $H_{\rm \em eff}$ writes as {\em three} (hermitian)
operators. The first one, associated to $\xi_{mnij}^{\rm dir}$, comes
from
direct Coulomb terms in which all interactions are unambiguously
interactions between excitons. The second operator, associated to
$\xi_{mnij}^{\rm exc}$, comes from exchange Coulomb terms in which
the
concept of interactions {\em between} excitons is rather ambiguous,
since
the e-h contributions are interactions {\em inside} one of the four
($m,n,i,j$) excitons. The third operator, associated to
$\eta_{mnij}$, is
purely fermionic. It appears in all non-diagonal scatterings. Let us
stress that, when properly symmetrized, the former ${\cal E}_{mnij}$
is
equal to $\xi_{mnij}^{\rm dir} - \xi_{mnij}^{\rm right}$ so that it
is
correct for ($m n$)=($i j$) only \cite{note10}.

iii) In the {\em three-exciton} subspace, we find that, except for
diagonal processes, Eq.~(\ref{eq13}) cannot be fulfilled with
two-body
scatterings only : Besides the terms found for $N=2$, the effective
hamiltonian must contain a purely fermionic 3-body operator,
\begin{equation}
H_{\rm \em eff}^{(3)} = \frac{1}{3!} \, \sum_{lmn,ijk} [-
\eta_{lmn,ijk} +
{\rm O}\big( (a_x^3/{\cal V})^{2} \big)] \, \overline{B}_{l}^{+} \,
\overline{B}_{m}^{+} \,\overline{B}_{n}^{+} \, \overline{B}_{i} \,
\overline{B}_{j} \, \overline{B}_{k} ,
\label{eq19}
\end{equation}
\begin{equation}
\eta_{lmn,ijk} = \frac{1}{3} \, (1- \delta_{lmnijk}) \, \Big\{ \big[
E_l
\, \delta_{l i} \, \lambda_{mnjk} + (i \leftrightarrow j) + (i
\leftrightarrow k)\big] + (l \leftrightarrow m) + (l \leftrightarrow
n)
\Big\} ,
\label{eq20}
\end{equation}
where $\delta_{lmnijk} =1$ if ($lmn$) = ($ijk$) and 0 otherwise.

In a similar way, Eq.~(\ref{eq13}) written in the $N$-exciton
subspace,
shows that $H_{\rm \em eff}$ must contain a set of (3,...,$N$)-body
operators of the order of $a_x^3/{\cal V}$ as the $X$-$X$ part. This
is
after all not surprising: Pauli exclusion being $N$-body by essence
since
{\em all} the electrons of the $N$ excitons must be different, the
fermionic character of the exciton has to appear through $N$-body
scatterings. Pauli exclusion between close-to-boson particles in fact
generates a new "many-body" effect which is conceptually quite
different
from the usual one. Indeed, Coulomb interaction being a 2-body
interaction, the usual many-body effects it induces are due to ($2
\times
2$) correlations between couples of electrons or holes. Here, Pauli
exclusion is $N$-body in itself, so that the many-body effects it
induces
are already in the bare operators.

The consequences of this work as well as its extension to excitons
with
angular momentum variables (easy to include along reference
\cite{MCPRB90}) will be presented in an extended paper.

\vfill
\eject

\appendix

\section{}

\begin{itemize}
\item The exact exciton creation operator $B_i^+$ is related to the
creation operators of free e-h pairs with same total momentum
${\mathbf
q}_i$ through
\begin{equation}
B_i^+ = \sum_{{\mathbf k}_i} \langle {\mathbf k}_i | x_{\nu_i}
\rangle \,
a_{{\mathbf K}_i^{e}}^+ \, b_{{\mathbf K}_i^{h}}^+ , \hskip2cm
a_{{\mathbf
K}_i^{e}}^+ \, b_{{\mathbf K}_i^{h}}^+ = \sum_{\nu_i} \langle
x_{\nu_i} |
{\mathbf k}_i \rangle \, B_i^+ ,
\label{eqa1}
\end{equation}
$| x_{\nu_i} \rangle$ being the exciton relative motion eigenstate,
and
${\mathbf K}_i^{e}={\mathbf k}_i+\alpha_e \, {\mathbf q}_i$,
${\mathbf
K}_i^{h} = -{\mathbf k}_i+\alpha_h \, {\mathbf q}_i$ the e and h
momenta
of the (${\mathbf k}_i,{\mathbf q}_i$) free pair, with $\alpha_{e,h}
=
m_{e,h}/(m_e+m_h)$.

\item Using these relations, we can write $V_i^+$ as
\cite{MCRC,MCReport,note12}
\begin{equation}
V_i^+ = \sum_{m,{\mathbf q} \neq {\mathbf 0}} V_{{\mathbf q}} \,
\gamma_{m
i}({\mathbf q}) \, B_m^+ \, \sum_{{\mathbf p}} (a_{{\mathbf
p}-{\mathbf
q}}^+ \, a_{{\mathbf p}} - b_{{\mathbf p}-{\mathbf q}}^+ \,
b_{{\mathbf
p}}) ,
\label{eqa2}
\end{equation}
where $V_{{\mathbf q}} = 4 \, \pi e^2/{\cal V} q^2$ in 3D.
$\displaystyle
\gamma_{m i}({\mathbf q}) = \delta_{{\mathbf q}_m,{\mathbf
q}_i+{\mathbf
q}} \, \langle x_{\nu_m} | e^{i \alpha_h \, {\mathbf q} \cdot
{\mathbf r}}
- e^{- i \alpha_e \, {\mathbf q} \cdot {\mathbf r}} |x_{\nu_i}
\rangle$
characterizes the scattering of an $i$ exciton into a $m$ state under
a
${\mathbf q}$ excitation. Inserting Eq.~(\ref{eqa2}) into
Eq.~(\ref{eq3}),
we find
\begin{equation}
\xi_{mnij}^{\rm dir} = \frac{1}{2} \, \sum_{{\mathbf q} \neq {\mathbf
0}}
V_{{\mathbf q}} \, \gamma_{m i}({\mathbf q}) \, \gamma_{n j}({\mathbf
-
q}) + (m \leftrightarrow n) ,
\label{eqa3}
\end{equation}
in which we have symmetrized $\xi_{mnij}^{\rm dir}$ in order to have
it
invariant under $(m \leftrightarrow n)$ or $(i \leftrightarrow j)$.
It
will be useful to note that $\xi_{mnij}^{\rm dir}$ also reads
\begin{eqnarray}
\xi_{mnij}^{\rm dir} & = & \sum_{{\mathbf k}_i,{\mathbf k}_j,{\mathbf
k}_m,{\mathbf k}_n} \,
\langle x_{\nu_m} | {\mathbf k}_m \rangle \, \langle x_{\nu_n} |
{\mathbf
k}_n \rangle \,
\langle {\mathbf k}_i | x_{\nu_i} \rangle \, \langle {\mathbf k}_j |
x_{\nu_j} \rangle \,
\Big[ \sum_{{\mathbf q} \neq {\mathbf 0}} V_{{\mathbf q}} \,
\delta_{mnij}^{\rm dir}({\mathbf q}) \Big] , \label{eqa4} \\
\delta_{mnij}^{\rm dir}({\mathbf q}) & = & \frac{1}{2} \, \Big(
\delta_{{\mathbf K}_m^{e},{\mathbf K}_i^{e}+{\mathbf q}} \,
\delta_{{\mathbf K}_m^{h},{\mathbf K}_i^{h}} - (e \leftrightarrow h)
\Big)
\, \Big( \delta_{{\mathbf K}_n^{e},{\mathbf K}_j^{e}-{\mathbf q}} \,
\delta_{{\mathbf K}_n^{h},{\mathbf K}_j^{h}} - (e \leftrightarrow h)
\Big)
+ (m \leftrightarrow n) . \label{eqa5}
\end{eqnarray}
$\delta_{mnij}^{\rm dir}({\mathbf q})$ corresponds to the set of
momentum
conservations of all direct interactions between two excitons having
a
${\mathbf q}$ momentum transfer. Since its four $\delta$'s
impose ${\mathbf q}_i + {\mathbf q}_j = {\mathbf q}_m + {\mathbf
q}_n$,
there are three conditions only between the four ${\mathbf k}$'s so
that
the sum of Eq.~(\ref{eqa4}) has only one free ${\mathbf k}$. We can
easily
check that Eq.~(\ref{eqa4}) reads, in real space,
\begin{eqnarray}
\xi_{mnij}^{\rm dir} & = & \frac{1}{2} \, \int de_1 de_2 dh_1 dh_2 \,
\phi_m^*(e_1,h_1) \, \phi_n^*(e_2,h_2) \big( V_{e_1 e_2} + V_{h_1
h_2} -
V_{e_1 h_2} - V_{e_2 h_1} \big) \nonumber\\
 & & \hskip2cm \times \, \phi_i(e_1,h_1) \, \phi_j(e_2,h_2) + (m
\leftrightarrow n) .
\label{eqa6}
\end{eqnarray}

\item Using Eq.~(\ref{eqa1}), we find that $\lambda_{mnij}$, defined
in
Eq.~(\ref{eq4}), reads as the expression (\ref{eqa4}) of
$\xi_{mnij}^{\rm
dir}$, except for the last bracket which is replaced by
\begin{equation}
\Delta_{mnij} = \frac{1}{2} \, \delta_{{\mathbf K}_m^{e},{\mathbf
K}_i^{e}} \, \delta_{{\mathbf K}_m^{h},{\mathbf K}_j^{h}} \,
\delta_{{\mathbf K}_n^{e},{\mathbf K}_j^{e}} \, \delta_{{\mathbf
K}_n^{h},{\mathbf K}_i^{h}} + (m \leftrightarrow n)
\label{eqa7}
\end{equation}
These $\delta$'s correspond to cross the e or the h of the two
excitons. In real space, $\lambda_{mnij}$ reads
\begin{equation}
\lambda_{mnij} = \frac{1}{2} \, \int de_1 de_2 dh_1 dh_2 \,
\phi_m^*(e_1,h_1) \, \phi_n^*(e_2,h_2) \, \phi_i(e_1,h_2) \,
\phi_j(e_2,h_1) + (m \leftrightarrow n) .
\label{eqa8}
\end{equation}

\item Eqs.~(\ref{eqa4}), (\ref{eqa5}), (\ref{eqa7}) show that
$\xi_{mnij}^{\rm right}$, defined in Eq.~(\ref{eq9}), reads as
$\xi_{mnij}^{\rm dir}$ (Eq.~(\ref{eqa4})) with $\delta_{mnij}^{\rm
dir}({\mathbf q})$ replaced by $\delta_{mnij}^{\rm right}({\mathbf
q})$
deduced from Eq.~(\ref{eqa5}) by (${\mathbf K}_i^{h} \leftrightarrow
{\mathbf K}_j^{h}$). The set of $\delta$'s of this
$\delta_{mnij}^{\rm
right}({\mathbf q})$ corresponds to the momentum conservations for
exchange
processes.

\end{itemize}

\vfill


\begin{references}
\bibitem{note9} See Eqs. (5.7-9) of Ref. \cite{Haug}. By using this
``unit-operator projection'' technique in a consistent way, it is
however
possible to produce the two missing terms ($V_{e_{1} h_{1}}+ V_{e_{2}
h_{2}}$) : One just has to form the ($m,n$) excitons with
($e_{1},h_{2}$)
and ($e_{2},h_{1}$) and the ($i,j$) excitons with ($e_{1},h_{1}$) and
($e_{2},h_{2}$). Let us stress that within this technique, based on
Coulomb interaction only, there is no way to produce the purely
fermionic
term $\eta_{mnij}$ in Eq.~(\ref{eq17}).
\bibitem{note10} While a correct diagonal term is enough to calculate
$\langle H_{sc} \rangle$, non-diagonal terms are necessary for
effects
like the ones considered in Ref. \cite{Inoue}.
\bibitem{note12} In Ref.~\cite{MCRC,MCReport}, $V_i^+$ was written in
terms of free e-h pairs instead of $B_i^+$'s, which was convenient
enough
for the purpose of these past works.
\end{references}
\end{document}